\documentclass[prx,twocolumn,floatfix,a4paper,superscriptaddress]{revtex4}

\usepackage{bm,color,amsmath,txfonts}
\usepackage{graphicx}
\usepackage{siunitx}
\usepackage{subfigure}
\usepackage{verbatim}
\usepackage{dcolumn}
\usepackage{bm}
\usepackage{epsf}
\usepackage{xcolor}
\usepackage{hyperref}
\usepackage{hhline}
\usepackage{float}
\usepackage{enumerate}
\usepackage{bbm}
\usepackage{lipsum}
\usepackage{mathrsfs}
\usepackage{ulem}

\begin{document}

\title{Entangling cavity-magnon polaritons by interacting with phonons}

\author{Xuan Zuo}
\affiliation{Zhejiang Key Laboratory of Micro-nano Quantum Chips and Quantum Control, School of Physics, and State Key Laboratory for Extreme Photonics and Instrumentation, Zhejiang University, Hangzhou 310027, China}
\author{Zhi-Yuan Fan}
\affiliation{Zhejiang Key Laboratory of Micro-nano Quantum Chips and Quantum Control, School of Physics, and State Key Laboratory for Extreme Photonics and Instrumentation, Zhejiang University, Hangzhou 310027, China}
\author{Hang Qian}
\affiliation{Zhejiang Key Laboratory of Micro-nano Quantum Chips and Quantum Control, School of Physics, and State Key Laboratory for Extreme Photonics and Instrumentation, Zhejiang University, Hangzhou 310027, China}
\author{Rui-Chang Shen}
\affiliation{Zhejiang Key Laboratory of Micro-nano Quantum Chips and Quantum Control, School of Physics, and State Key Laboratory for Extreme Photonics and Instrumentation, Zhejiang University, Hangzhou 310027, China}
\author{Jie Li}\thanks{jieli007@zju.edu.cn}
\affiliation{Zhejiang Key Laboratory of Micro-nano Quantum Chips and Quantum Control, School of Physics, and State Key Laboratory for Extreme Photonics and Instrumentation, Zhejiang University, Hangzhou 310027, China}

\begin{abstract}
We show how to entangle two cavity-magnon polaritons (CMPs) formed by two strongly coupled microwave cavity and magnon modes. This is realized by introducing vibration phonons, via magnetostriction, into the system that are dispersively coupled to the magnon mode. Stationary entanglement between two CMPs can be achieved when they are respectively resonant with the two sidebands of the drive field scattered by the phonons, and when the proportions of the cavity and magnon modes in the two polaritons are appropriately chosen. The entangled CMPs are macroscopic quantum states as the magnon mode contains a large number of spins, and can lead to the emission of frequency-entangled microwave photons, which find broad applications in microwave quantum information processing and quantum metrology.
\end{abstract}

\maketitle

\section{Introduction}

Cavity magnonics~\cite{Bauer}, which studies the interaction between microwave cavity photons and magnons in magnetic materials, e.g., yttrium iron garnet (YIG), has attracted great attention and made significant progress in the past decade. The development of the field has been greatly accelerated since the strong cavity-magnon coupling was experimentally achieved~\cite{Huebl,Naka14,Tang14}, as theoretically predicted in Ref.~\cite{Flatte}, benefiting from many excellent properties of the YIG, such as a high spin density and a low dissipation rate. The strong coupling leads to the hybridization of the cavity and magnon modes, forming two cavity-magnon polaritons (CMPs). The two hybridized modes are not entangled due to the linear excitation-exchange (beam-splitter-type) interaction between the cavity and magnon modes~\cite{Zuo24}, as the generation of entanglement typically requires the parametric process, external quantum fields, or other novel entangling mechanisms. This is the fundamental reason why it is difficult to entangle two CMPs. Although many proposals have been put forward in the field for preparing entangled states~\cite{Jie18,Jie19,GSA19,Tan19,Yuan20,Yuan20b,Li20,Jaya,YM,Zheng21,Luo,Simon,Chen21,Kong,Wu21,Yang21,Mou,Xie,JingJ,Ren,Hussain22,Qiu22,Zhou22,Qian23,Chen23,Xie23,Yang24,Tan23}, to date a theory for entangling two CMPs is still lacking.

Here, we offer a route to solve this fundamental problem. We show that by introducing a phonon mode that is {\it dispersively} coupled to the magnon mode (thus forming the cavity magnomechanical system~\cite{Zuo24}), the two CMPs can be entangled in a stationary state. {The phonon mode introduces a parametric down-conversion (PDC) interaction associated with the lower-frequency polariton, while a state-swap interaction related to the higher-frequency polariton.} By appropriately adjusting the strengths of the PDC and the state-swap interactions, the two CMPs get entangled via the mediation of phonons. The entanglement is robust against thermal noises and dissipation rates of the three modes. 

The paper is structured as follows. In Sec.~\ref{model}, we introduce the model and provide the Hamiltonian and Langevin equations of the system. We further derive the linearized Langevin equations for the quantum fluctuations of the system and show how to obtain the steady-state solutions. In Sec.~\ref{resul}, we explain the underlying mechanism for achieving the steady-state entanglement of the two CMPs and of the associated two cavity output modes, and present the main results. We then provide optimal conditions for the entanglement and study the effects of the dissipations and input noises of the system on the entanglement. Finally, we summarize our findings in Sec.~\ref{conc}.

\begin{figure}[b]
	\includegraphics[width=\linewidth]{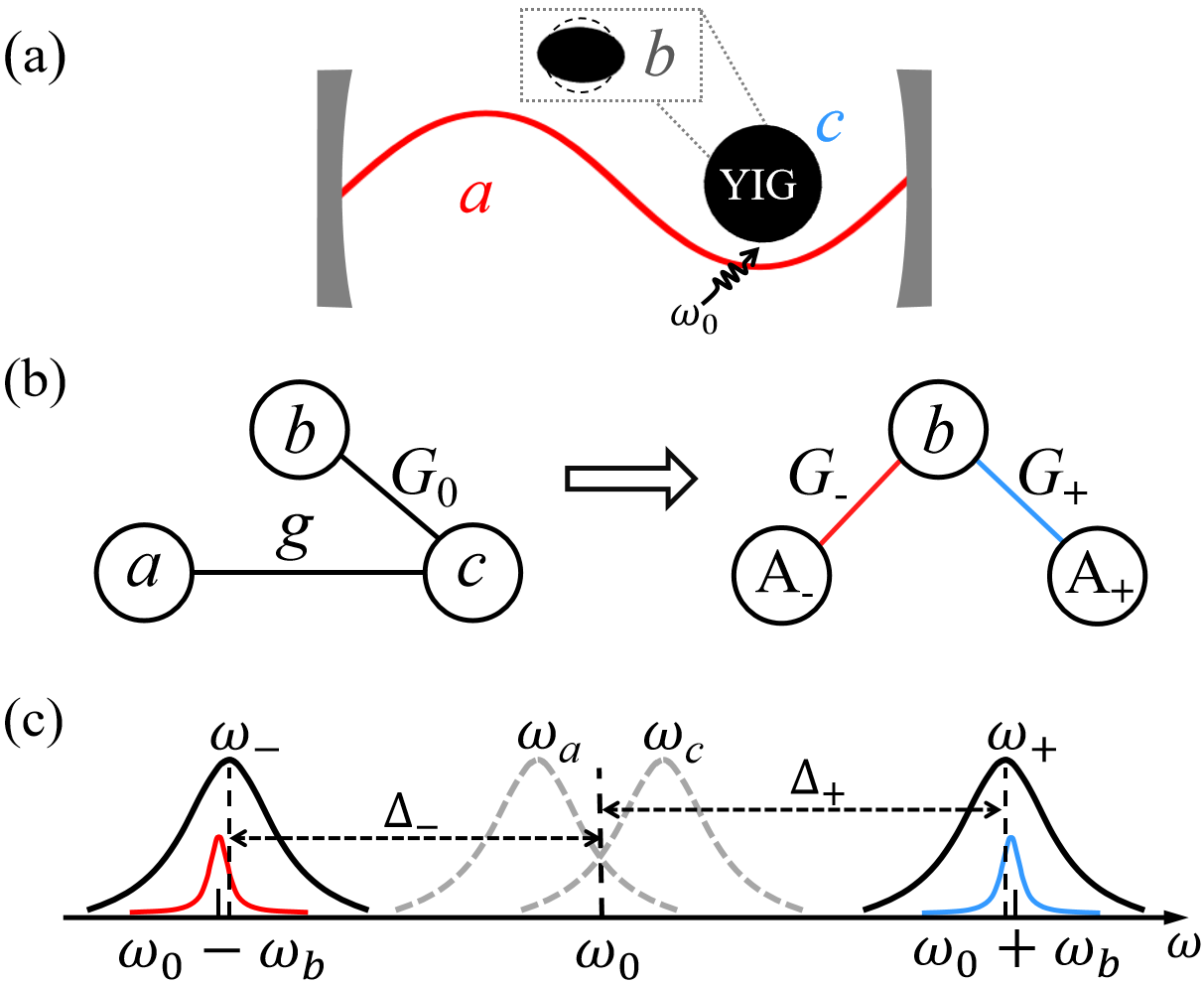}
	\caption{{(a) Sketch of the system: A YIG sphere, supporting both a magnon mode ($c$) and a phonon mode ($b$), is placed inside a microwave cavity ($a$). The magnon mode is directly driven by a microwave field with frequency $\omega_0$.} (b) The cavity magnomechanical system in the strong cavity-magnon coupling regime. The magnon mode couples to the cavity mode via a beam-splitter interaction and to the phonon mode via a dispersive interaction. The cavity and magnon modes are strongly coupled forming two CMP modes $A_-$ and $A_+$, of which both are coupled to the phonon mode. (c) Frequencies and linewidths of the system. The magnon mode at frequency $\omega_c$ is strongly driven by a microwave field at frequency $\omega_0$, and the phonons at frequency $\omega_b$ scatters the driving photons onto the Stokes sideband at $\omega_0-\omega_b$ and the anti-Stokes sideband at $\omega_0+\omega_b$. When the frequencies of two CMPs match the two sidebands, two CMPs are entangled via the mediation of phonons.}
	\label{fig1}
\end{figure}

\section{The model}\label{model}

{The cavity magnomechanical system consists of a microwave cavity mode, a magnon mode, and a mechanical vibration mode~\cite{Zuo24,Tang,Davis,Jie22}, as shown in Fig.~\ref{fig1}(a). The magnon mode is a spin wave, i.e., the collective motion of a large number of spins in magnetic materials, e.g., a YIG sphere.} The cavity and magnon modes are coupled via the magnetic dipole interaction, and the coupling strength can be much larger than the cavity and magnon dissipation rates, leading to two CMPs. The strong coupling results in the normal-mode splitting, as experimentally observed in the cavity magnonic systems~\cite{Huebl,Naka14,Tang14}.  {The magnon mode further dispersively couples to a deformation phonon mode of the YIG sphere via the magnetostrictive interaction.} This is typically the situation where the resonance frequency of the phonon mode is much lower than that of the magnon mode, e.g., for large-size YIG spheres~\cite{Zuo24}. The system then becomes a hybrid three-mode system, as depicted in Fig.~\ref{fig1}(b), with the Hamiltonian given by
\begin{align}\label{HHH}
	\begin{split}
		H/\hbar &= \! \omega_{a} a^\dagger a + \omega_{c} c^\dagger c + \omega_b b^\dagger b + G_0 c^\dagger c \left( b + b^\dagger \right)		
		\\& + g \left(a^\dagger c+ac^\dagger \right) + i\Omega \left(c^\dagger e^{-i\omega_0t}-c e^{i\omega_0t} \right) ,
	\end{split}
\end{align}
where $j$ ($j^\dagger$), $j=a,c,b$, are the annihilation (creation) operators of the cavity, magnon and phonon modes, respectively, satisfying the commutation relation $[j,j^{\dag}]=1$. $\omega_j$ are their corresponding resonance frequencies, and $\omega_b \ll  \omega_a, \omega_c$. $g$ denotes the linear cavity-magnon coupling strength, and
$G_0$ is the single-magnon magnomechanical coupling strength.
Due to the dispersive nature, $G_0$ is typically weak, but the effective magnomechanical coupling can be significantly improved by driving the magnon mode with a strong microwave field (at frequency $\omega_0$ and amplitude $B_0$), and the corresponding Rabi frequency $\Omega = \frac{\sqrt{5}}{4} \gamma_0 \sqrt{N} B_0$~\cite{Jie18}, where $\gamma_0/2\pi = 28$ GHz T$^{-1}$ is the gyromagnetic ratio and $N = \rho V$ is the total number of spins, with $\rho = 4.22 \times 10^{27}$ m$^{-3}$ being the spin density of the YIG and $V$ as the volume of the YIG sphere.

Since the strongly coupled cavity and magnon modes are hybridized, it is more convenient to describe the cavity-magnon system with two polariton operators $A_{\pm}$. The Hamiltonian~\eqref{HHH} can then be rewritten in terms of  $A_{\pm}$ and, in the interaction picture with respect to $\hbar \omega_0 (A_+^\dagger A_+ + A_-^\dagger A_-)$, is given by (Appendix A)
\begin{align}
	\begin{split}
		H/\hbar &= \! \Delta_{+} A_+^\dagger A_+ + \Delta_{-} A_-^\dagger A_- + \omega_b b^\dagger b  + G_0 \left( b + b^\dagger \right)
		\\& \times \left[A_+^\dagger A_+ \sin^2\theta + A_-^\dagger A_- \cos^2\theta  +  \frac{1}{2} \left(A_+^\dagger A_-  + A_-^\dagger A_+ \right) \sin2\theta \right]
		\\&+ i\Omega \left(A_+^\dagger \sin\theta + A_-^\dagger \cos\theta - A_+ \sin\theta - A_- \cos\theta \right) ,
	\end{split}
\end{align}
where $A_{+}$ and $A_{-}$ are the annihilation operators of the two CMPs, which are the hybridization of the cavity and magnon modes via $A_+=a \cos\theta + c\sin\theta$ and $A_-=- a\sin\theta + c\cos\theta$, with $ \theta =\frac{1}{2} \arctan \frac{2g}{\omega_a-\omega_c}$.  $A_{\pm}$ satisfy the bosonic commutation relation $[k,k^\dagger]=1$ ($k=A_{\pm}$). $\Delta_{\pm} = \omega_{\pm}-\omega_0$ denote the polariton-drive detunings, where $\omega_{\pm}=\frac{1}{2}\Big[\omega_a+\omega_c \pm \sqrt{(\omega_a-\omega_c)^2+4g^2} \Big]$ are the frequencies of the two CMPs, cf. Fig.~\ref{fig1}(c).

By including the dissipation and input noise of each mode, we obtain the following quantum Langevin equations (QLEs) of the system (Appendix B):
\begin{align}\label{QLEAA}
	\begin{split}
		\dot{A}_+=&-i\Delta_+ A_+ - i G_0 \left( b + b^\dagger \right) \left( A_+ \sin^2\theta + A_-\sin \theta \cos \theta \right) 
		\\& - \kappa_+ A_+ - \delta\kappa A_-  + \Omega \sin\theta + \sqrt[]{2\kappa_+}A_+^{in},  \\
		\dot{A}_-=&-i\Delta_- A_- - i G_0 \left( b + b^\dagger \right) \left(A_- \cos^2\theta + A_+\cos \theta \sin \theta \right) 
		\\& - \kappa_- A_- - \delta\kappa A_+ + \Omega \cos\theta + \sqrt[]{2\kappa_-}A_-^{in}, \\
		\dot{b}=&-i\omega_b b - i G_0 \Big(A_+^\dagger A_+ \sin^2\theta+A_+^\dagger A_- \sin\theta \cos\theta 
		\\&+ A_-^\dagger A_+ \cos\theta \sin\theta + A_-^\dagger A_- \cos^2\theta \Big) - \kappa_b b + \sqrt[]{2\kappa_b} b^{in},
	\end{split}
\end{align}
where $\kappa_+ \equiv  \kappa_a \cos^2\theta + \kappa_c \sin^2\theta $ and $\kappa_- \equiv \kappa_a \sin^2\theta + \kappa_c \cos^2\theta $ are the dissipation rates of the two polaritons $A_+$ and $A_-$, respectively, and $\delta\kappa \equiv (\kappa_c-\kappa_a) \sin\theta \cos\theta$ denotes the coupling strength between the two CMPs due to the unbalanced dissipation rates $\kappa_a \neq \kappa_c$, with $\kappa_j$ being the dissipation rate of the mode $j$ ($j=a,c,b$). $A_+^{in} \equiv (\sqrt[]{2\kappa_a} \cos \theta a^{in} + \sqrt[]{2\kappa_c} \sin \theta c^{in}) /\sqrt[]{2\kappa_+}$ and $A_-^{in} \equiv (- \sqrt[]{2\kappa_a} \sin \theta a^{in} + \sqrt[]{2\kappa_c} \cos \theta c^{in}) /\sqrt[]{2\kappa_-}$ represent the noises entering the two CMPs, which are related to the input noises $a^{in}$, $c^{in}$ of the original cavity and magnon modes, and 
$b^{in}$ is the input noise of the phonon mode. The input noises $j^{in}(t)$, $j\,\,{=}\,\,a,c,b$, are zero-mean and characterized by the correlation functions~\cite{zoller}: $\langle j^{in}(t)j^{in\dagger}(t^\prime) \rangle=[N_j(\omega_j)+1]\delta(t-t^\prime)$, $\langle j^{in\dagger}(t)j^{in}(t^\prime) \rangle=N_j(\omega_j)\delta(t-t^\prime)$, with $N_j(\omega_j)=[\exp[(\hbar \omega_j/k_BT)]-1]^{-1}$ being the equilibrium mean thermal excitation number of the mode $j$, and $T$ as the bath temperature.

Since the magnon mode is strongly driven and due to its excitation-exchange interaction with the cavity mode, the two CMPs have large amplitudes $| \langle A_+ \rangle|,\, | \langle A_- \rangle| \gg 1$ at the steady state. This allows us to linearize the nonlinear dispersive interaction around the large average values~\cite{DV,Jie18}. This is implemented by writing each mode operator $O$, $O=A_+,A_-,b$, as the sum of its classical average and quantum fluctuation operator, i.e., $O=\langle O \rangle + \delta O$, and neglecting small second-order fluctuation terms in Eq.~\eqref{QLEAA}.  We aim to study quantum entanglement between the two CMPs, and hence focus on the dynamics of the quantum fluctuations. The quantum fluctuations of the system $(\delta A_+,\delta A_-, \delta b)$ are governed by the following linearized QLEs (Appendix B):
\begin{align}\label{QLEAAfluc}
	\begin{split}
		\dot{\delta A_+}{=}&\,{-} \big( i{\Delta}_+ + \kappa_+ \big) \delta A_+ {-}\, \delta\kappa \delta A_- {-} \,G_{+b} \frac{\delta b + \delta b^\dagger}{2} {+}\, \sqrt[]{2\kappa_+}A_+^{in},  \\
		\dot{\delta A_-}{=}&\,{-} \big( i{\Delta}_- + \kappa_- \big) \delta A_- {-} \, \delta\kappa  \delta A_+ {-} \,G_{-b} \frac{\delta b + \delta b^\dagger}{2} {+}\, \sqrt[]{2\kappa_-}A_-^{in},  \\
		\dot{\delta b}\,{=}& \,{-} \big( i \omega_b + \kappa_b \big) \delta b  \,{-} { \left(\frac{G_{+b}}{2}  \delta A_+^\dagger + \frac{G_{-b}}{2}  \delta A_-^\dagger - \rm{H.c.}\right) } + \sqrt[]{2\kappa_b}b^{in},
	\end{split}
\end{align}
where $G_{+b} \equiv G_{+-} \sin \theta$ ($G_{-b}  \equiv G_{+-} \cos \theta$) represents the coupling strength between the polariton $A_+$ ($A_-$) and the phonon mode $b$, with $G_{+-} \equiv G_+ \sin\theta + G_- \cos \theta$, and $G_{\pm}= i 2 G_0\langle A_{\pm} \rangle$ being the enhanced dispersive coupling strengths associated with the two CMPs.  In obtaining Eq.~\eqref{QLEAAfluc}, we neglect the linear coupling terms between the two CMPs ${\cal G} \left(\delta A_+^\dag \delta A_-+ \delta A_+ \delta A_-^\dag \right)$, where ${\cal G} = G_0 {\rm Re} \langle b \rangle \sin 2\theta$, due to their weak strength and negligible impact on the entanglement.   Under the optimal conditions for the entanglement $|{\Delta}_{\pm}| \simeq \omega_b \gg \kappa_{\pm}$ (cf. Fig.~\ref{fig1}(c)), as will be discussed later, we obtain approximate analytical expressions of the steady-state averages, i.e.,
\begin{align}\label{stSol}
	\begin{split}
		\langle A_+ \rangle &\simeq \frac{\delta\kappa \Omega \cos \theta - i \Omega \sin \theta({\Delta}_- - i \kappa_-)}{({\Delta}_- - i \kappa_-)({\Delta}_+ - i \kappa_+) + \delta\kappa^2}, \\	 \langle A_- \rangle &\simeq \frac{\delta\kappa \Omega \sin \theta - i \Omega \cos \theta({\Delta}_+ - i \kappa_+)}{({\Delta}_- - i \kappa_-)({\Delta}_+ - i \kappa_+) + \delta\kappa^2},\\{\rm Re} \langle b \rangle&=-\frac{G_0}{\omega_b} \big| \langle A_+ \rangle \sin \theta + \langle A_- \rangle \cos \theta \big|^2.
	\end{split}
\end{align}
Because of the weak coupling $G_0$, the frequency shift caused by the dispersive coupling is typically much smaller than the resonance frequency $\omega_b$, as observed in the magnomechanical experiments~\cite{Tang,Davis,Jie22}. Therefore, in deriving Eqs.~\eqref{QLEAAfluc} and \eqref{stSol} we safely neglect this small frequency shift in the detunings $|{\Delta}_{\pm}| \simeq \omega_b$.

 The QLEs \eqref{QLEAAfluc} can be expressed using quadratures $(\delta X_{\pm}, \delta Y_{\pm}, \delta X_b, \delta Y_b)$, with $\delta X_{\pm}=(\delta A_{\pm} +\delta A_{\pm}^\dagger)/\!\sqrt{2}$, $\delta Y_{\pm}=i(\delta A_{\pm}^\dagger - \delta A_{\pm})/\!\sqrt{2}$, and $\delta X_b=(\delta b + \delta b^\dagger)/\!\sqrt{2}$, $\delta Y_b=i (\delta b^\dagger - \delta b)/\!\sqrt{2}$, and cast in the matrix form
\begin{align}\label{uRn}
	\begin{split}
	\dot{u}(t)={\cal R}u(t) + n(t),
	\end{split}
\end{align}
where $u(t)=[\delta X_+(t),\delta Y_+(t),\delta X_-(t),\delta Y_-(t),\delta X_b(t),\delta Y_b(t)]^{\rm T}$, $n(t)=[\sqrt[]{2\kappa_+}X_+^{in},\sqrt[]{2\kappa_+}Y_+^{in},\sqrt[]{2\kappa_-}X_-^{in},\sqrt[]{2\kappa_-}Y_-^{in},\sqrt[]{2\kappa_b}X_b^{in},\sqrt[]{2\kappa_b}Y_b^{in}]^{\rm T}$, and the drift matrix ${\cal R}$ is given by 
\begin{align}\label{RRR}
	\cal R=\begin{pmatrix}
		-\kappa_+ & {\Delta}_+ & -\delta\kappa & 0 & - {\rm Re}\,G_{+b} &0 \\
		-{\Delta}_+ & -\kappa_+ &0 & -\delta\kappa& - {\rm Im}\,G_{+b} & 0\\
		-\delta\kappa & 0 & -\kappa_- & {\Delta}_- & - {\rm Re}\,G_{-b} &0 \\
		0 & -\delta\kappa&-{\Delta}_- & -\kappa_- & - {\rm Im}\,G_{-b} & 0\\
		0 & 0 & 0 & 0 & -\kappa_b & \omega_b \\
		{- {\rm Im}\,G_{+b}} &{\rm Re}\,G_{+b} & { - {\rm Im}\,G_{-b}} & {\rm Re}\,G_{-b} & -\omega_b & -\kappa_b\\
	\end{pmatrix}.
\end{align} 
Since the quantum noises are Gaussian and the system dynamics is linearized, the steady state of the quadrature fluctuations is a continuous-variable three-mode Gaussian state, which can be completely characterized by a $6\times6$ covariance matrix (CM) $V$ with its entries defined as $V_{ij}=\frac{1}{2} \langle u_i(t)u_j({t}) + u_j({t})u_i(t) \rangle$ $(i,j=1,2,...,6)$.  The steady-state CM $V$ can be achieved by directly solving the Lyapunov equation~\cite{DV}
\begin{align}
	\begin{split}
	{\cal R} V+V{\cal R}^T = -D,
	\end{split}
\end{align}
where $D=\mathrm{Diag} \big[\kappa_+(2N_+ {+}\,1),\kappa_+(2N_+ {+}\,1),\kappa_-(2N_- {+}\,1),\kappa_-(2N_-$ $+1),\kappa_b(2N_b+1),\kappa_b(2N_b+1)\big]$
$+\,\frac{1}{2} \tan2\theta\,[-\kappa_+(2N_+ {+}\,1)+\kappa_-(2N_- {+}\,1)]\, \bm{\sigma}^x\,\otimes\,\rm{I}_{2\times2}\,\oplus\,\bm{0}_{2\times2}$ is the diffusion matrix, which is defined via $D_{ij}\,\delta(t \,{-} \,t')=\langle n_i(t)n_j(t')+n_j(t')n_i(t) \rangle/2$.  $\bm{\sigma}^x$ is the $x$-Pauli matrix and the mean thermal excitation numbers $N_{\pm}$ are related to $N_a$ and $N_c$ via $N_+ = \Big\{ \big[\kappa_a \cos^2 \theta (2 N_a + 1) + \kappa_c \sin^2 \theta (2 N_c + 1) \big]/\kappa_+ -1 \Big\}/2 $ and $ N_- = \Big\{ \big[\kappa_a \sin^2 \theta (2 N_a + 1) + \kappa_c \cos^2 \theta (2 N_c + 1)\big]/\kappa_- - 1 \Big\}/2 $.

{The two CMPs are inside the microwave cavity and their entanglement is in principle inaccessible, which, however, can be verified by measuring the entanglement of the corresponding two cavity output modes. In what follows, we show how to define the output modes and calculate the entanglement between them.} 
The output field of the microwave cavity can be derived from the intracavity field via the input-output relation: 
\begin{align}
	\begin{split}
	\delta a^{\rm out}(t) = \sqrt{2 \kappa_a} \delta a(t) - a^{\rm in}(t),
	\end{split}
\end{align}
which forms a {\it continuous} spectrum but can be filtered to define a set of independent output modes~\cite{Genes08}. Here, we focus on the output field around the Stokes and anti-Stokes sidebands, which share entanglement as will be discussed in Sec.~\ref{resul}. Specifically, two independent and well-resolved output modes are defined with the filter central frequencies $\Omega_1$ and $\Omega_2$, respectively, and the same frequency bandwidth $\sim {\tau}^{-1}$, which serve as controllable parameters in the filter function {(see Appendix C for more details).}

The steady state of the magnon, mechanical, and two output modes is fully characterized by the corresponding $8\times8$ CM, whose entries are defined as
\begin{align}
	\begin{split}
	V_{ij}^{\rm out} = \frac{1}{2} \big\langle u_i^{\rm out}(t)u_j^{\rm out}(t) + u_j^{\rm out}(t)u_i^{\rm out}(t) \big\rangle,
	\end{split}
\end{align}
where
\begin{align}
	\begin{split}
	u^{\rm out}(t)=\big[&\delta X_c(t), \delta Y_c(t), \delta X_b(t), \delta Y_b(t),\\ &\delta X_1^{\rm out}(t), \delta Y_1^{\rm out}(t), \delta X_2^{\rm out}(t), \delta Y_2^{\rm out}(t) \big]^{\rm T}
	\end{split}
\end{align}
is the vector of the quadrature fluctuations of the four modes. Taking the Fourier transform and utilizing the correlation functions of the input noises, one obtains {(Appendix C)}
\begin{align}
	\begin{split}
	V^{\rm out} =& \int  d\omega \tilde{T}(\omega) \left( \tilde{M}^{\rm ext}(\omega) + \frac{P^{\rm out}}{2 \kappa_a} \right)\\ \times &\,\, D^{\rm ext} \left( \tilde{M}^{\rm ext}(\omega)^\dagger + \frac{P^{\rm out}}{2 \kappa_a} \right) \tilde{T}(\omega)^\dagger,
	\end{split}
\end{align}
{where the expressions of the matrices $\tilde{T}$, $\tilde{M}^{\rm ext}$, $P^{\rm out}$, and $D^{\rm ext}$ are too lengthy to be presented in the main text, but are provided in the Appendix.}

We adopt the logarithmic negativity $E_N$~\cite{ln1,ln2,ln3} to quantify the entanglement between the two CMPs and between the two output modes, which is defined based on the $4\times4$ CM $V_4$ of the two polariton modes, or the two output modes ($V_4$ is extracted by removing irrelevant rows and columns in $V$ or $V^{\rm out}$). Specifically, for the two CMPs $E_N=\textup{max}\left[ 0,-\textup{ln}(2\eta^{-})\right]$, where $\eta^{-} \,\,{\equiv}\,\, 2^{-1/2}\, \big[ \Sigma \,\,{-}\,\, \big(\Sigma^2 \,\,{-}\,\, 4\, \textup{det}\, V_{4} \big)^{1/2}\big]^{1/2}$, and $V_{4}=$ $\big[V_+, V_{+-}; V_{+-}^{\rm T}, V_- \big]$, with $V_+, V_-$ and $V_{+-}$ being the $2\times2$ blocks of $V_{4}$, and $\Sigma \equiv \textup{det}\,V_++\textup{det}\, V_--2\textup{det}\,V_{+-}$.  Similarly, one can compute the logarithmic negativity of the two output modes.

 \section{Entanglement of two CMPs}
 \label{resul}

The mechanism of creating entanglement between the two CMPs is as follows. The phonons scatter the driving microwave photons at frequency $\omega_0$ onto two sidebands at $\omega_0 \pm \omega_b$ (cf. Fig.~\ref{fig1}(c)). When the frequencies of the two CMPs are adjusted to be resonant with the two sidebands, i.e., ${\Delta}_+=-{\Delta}_- \simeq \omega_b$, both the Stokes and anti-Stokes scatterings are effectively activated, where the Stokes scattering corresponds to the PDC interaction causing the lower-frequency polariton to be entangled with the phonon mode, while the anti-Stokes scattering leads to the state-swap (beam-splitter) interaction between the higher-frequency polariton and the phonon mode. Therefore, the two CMPs get entangled via the mediation of phonons when the above two processes are simultaneously activated.

\begin{figure}[t]
	\includegraphics[width=\linewidth]{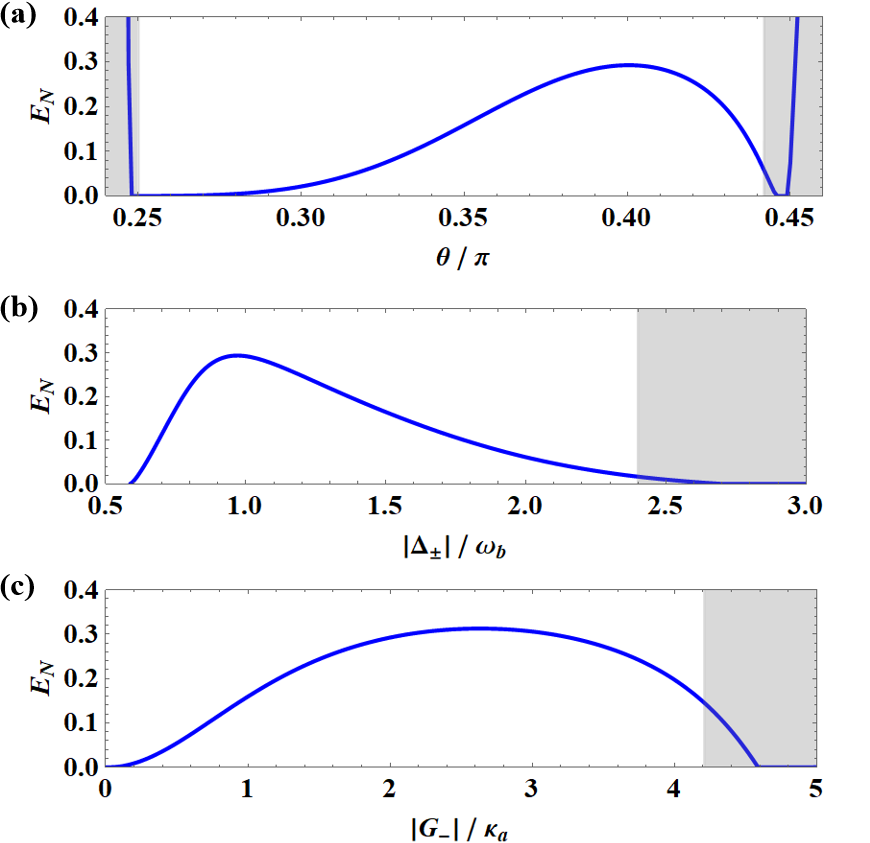}
	\caption{Stationary entanglement between two CMPs $E_N$ versus (a) $\theta$; (b) the polariton-drive detuning $|\Delta_{\pm}|$; (c) the coupling strength $|G_{-}|$. We take $|G_-|\,/\,2\pi = 2$  MHz and ${\Delta}_+=-{\Delta}_-=\omega_b$ in (a). The parameters of (b) [(c)], apart from $\Delta_{\pm}$ ($|G_{-}|$), correspond to those yielding an optimal angle $\theta \simeq {0.40}\pi$ in (a).  The grey areas denote the regimes where the system is unstable. See text for other parameters.}
	\label{fig2}
\end{figure}

The interactions between the CMPs and the phonon mode essentially result from the dispersive coupling between the phonon mode and the component of the magnon mode in the CMPs.  Therefore, by varying the proportion of the magnon mode in the polaritons (via altering $\theta$), one can adjust the {\it effective} strength of the PDC (state-swap) interaction associated with the Stokes (anti-Stokes) scattering. For the special case $\theta = \frac{\pi}{4}$, $\theta \in [0, \frac{\pi}{2}]$, where both the CMPs $A_{\pm}$ have equally weighted cavity and magnon modes, the strengths of the Stokes and anti-Stokes scatterings are equal. This point easily causes the system to be unstable (cf. Fig.~\ref{fig2}(a)) and only nonstationary entanglement could be produced~\cite{Jie15}. Therefore, a larger $\theta > \frac{\pi}{4}$ should be considered to obtain stationary entanglement, where the anti-Stokes scattering (for mechanical cooling) outperforms the Stokes scattering (for mechanical amplification). For a relatively small value of $\theta$, e.g., $\theta <\frac{\pi}{3}$, we find that the entanglement between the two CMPs is small, while the entanglement between the lower-frequency polariton $A_-$ and phonons is strong.  This indicates that the entanglement ($A_-$ \& $b$) is effectively generated by the Stokes scattering, but is not yet efficiently transferred to the higher-frequency polariton $A_+$. The state-swap interaction between the polariton $A_+$ and the phonon mode $b$ in the anti-Stokes scattering should thus be enhanced.  
To this end, we further increase $\theta$ to raise the proportion of the magnon mode in the polariton $A_+$, which enhances the strength of the anti-Stokes scattering.  Consequently, we see an efficient entanglement transfer and the two CMPs are strongly entangled around $\theta \simeq {0.40}\pi$, as clearly shown in Fig.~\ref{fig2}(a). However, $\theta$ cannot be too large (i.e., too close to $\frac{\pi}{2}$), as this reduces the proportion of the magnon mode in the polariton $A_-$ and thus the strength of the associated Stokes scattering, from which the entanglement of the system originates.  Therefore, an optimal $\theta$ exists for the entanglement as the result of the trade-off between the strengths of the Stokes and anti-Stokes scatterings, as seen in Fig.~\ref{fig2}(a).

In Fig.~\ref{fig2}, we use experimentally feasible parameters~\cite{Zuo24}: $\omega_a/2\pi=10$ GHz, $\omega_b/2\pi=10$ MHz, $\kappa_a/2\pi=\kappa_c/2\pi=1$ MHz, $\kappa_b/2\pi=100$ Hz, and at low temperature $T=10$ mK. We further consider the optimal detunings ${\Delta}_+=-{\Delta}_-=\omega_b$ in Fig.~\ref{fig2}(a), as analyzed above. For a given $\theta$, $g$ and $\omega_c$ can be determined by solving the equations $\theta=\frac{1}{2} \arctan \frac{2g}{\omega_a-\omega_c}$ and $ \sqrt{(\omega_a-\omega_c)^2+4g^2}=2\omega_b$ ($\omega_{a,b}$ are assumed constant). We note that the magnon frequency $\omega_c$ and the cavity-magnon coupling strength $g$ can be readily adjusted, respectively, by varying the bias magnetic field and the position of the YIG sphere inside the microwave cavity~\cite{Huebl,Naka14,Tang14}.  
The stability of the system is guaranteed by the negative eigenvalues (real parts) of the drift matrix ${\cal R}$. We find that, apart from the unstable region when $\theta$ is small ($\theta \le 0.25 \pi$),  the system can also be unstable when $\theta$ is too large. This is because the coupling strength $|G_{+}|$ (associated with the polariton $A_+$) rapidly increases as $\theta \to \frac{\pi}{2}$, due to the relation $|\frac{G_+}{G_-}| \simeq \tan \theta$ and the fact that $|G_{-}|$ is fixed in Fig.~\ref{fig2}(a).

An optimal $\theta \simeq {0.40}\pi$ gives the maximum entanglement in Fig.~\ref{fig2}(a), which corresponds to $g/2\pi \simeq {5.88}$ MHz and $\omega_c/2\pi \simeq {10.0162}$ GHz. To investigate the effect of the deviation of the polaritons from the mechanical sidebands, we change $\omega_c$ in Fig.~\ref{fig2}(b) to vary the polariton frequencies (i.e., $\Delta_{\pm}$).   Specifically, we set the drive frequency $\omega_0=\frac{\omega_a + \omega_c}{2}$, which ensures two symmetric CMPs ($\Delta_+=|\Delta_-|$) and mechanical sidebands with respect to the drive frequency.
It shows that when the two CMPs deviate from the two mechanical sidebands ($|{\Delta}_{\pm}|$ away from $\omega_b$), the entanglement reduces, confirming our earlier analysis on the optimal condition  ${\Delta}_+=-{\Delta}_- \simeq \omega_b$ for the entanglement.  
The unstable region is because that a large $|{\Delta}_{\pm}|$ corresponds to $\theta$ approaching $\frac{\pi}{2}$, and thus a large value of $|G_{+}|$. This is similar to the reason of the instability on the right side of Fig.~\ref{fig2}(a).

In Fig.~\ref{fig2}(c), we plot the entanglement versus the effective coupling rate $G_{-}$. Clearly, as $| G_{-}|$ grows the entanglement reaches its maximum in the stable regime and then reduces before entering the unstable regime. This means that one can obtain the maximum entanglement {\it in the stable regime}, and the maximum entanglement is no longer constrained by the stability condition~\cite{DV,KH}. This implies that the entanglement mechanism presented here fundamentally differs from those in the protocols of, e.g., Refs.~\cite{Jie18,DV}.

\begin{figure}[t]
	\includegraphics[width=\linewidth]{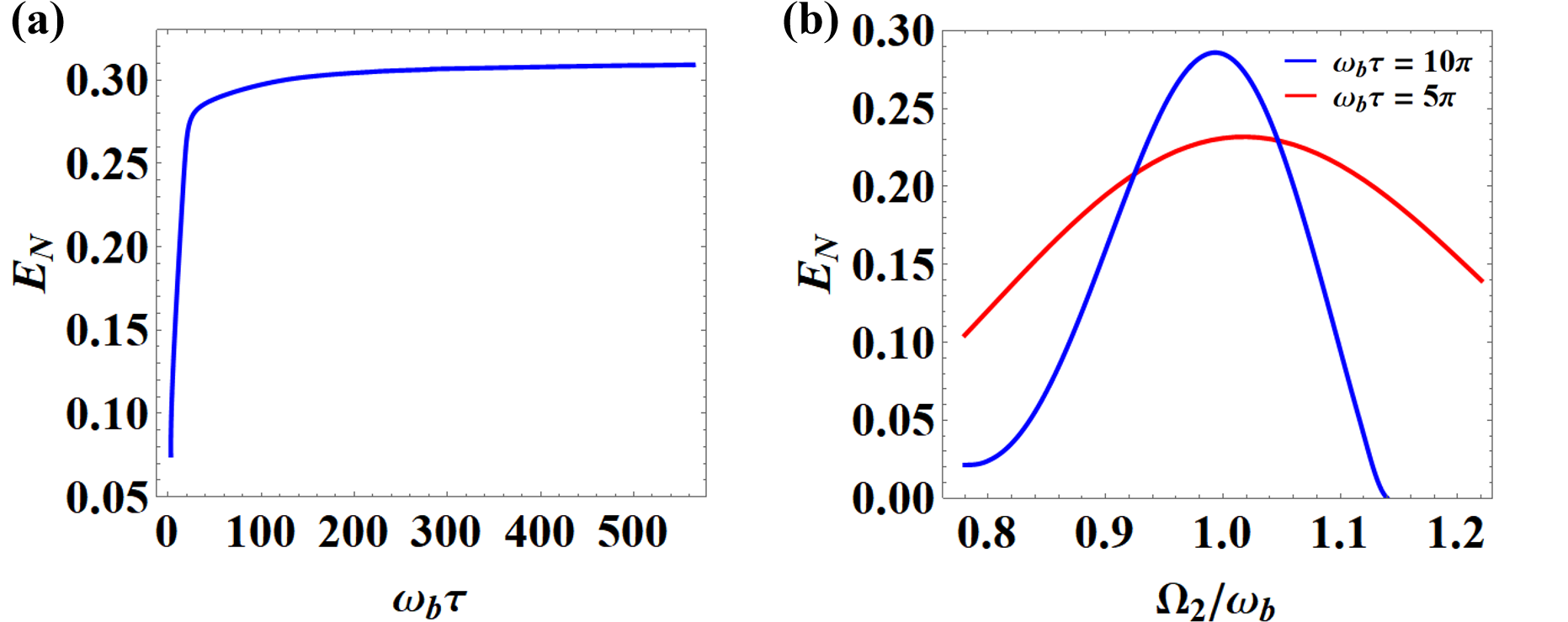}
	\caption{Stationary entanglement $E_N$ of two filtered output modes versus (a) inverse bandwidth $\omega_b \tau$; (b) central frequency of the second output mode $\Omega_2$ with different inverse bandwidths: $\omega_b \tau = 5 \pi$  and $10 \pi$. We take $\Omega_1 = -\Omega_2 = -\omega_b$ in (a), $\Omega_1 = -\omega_b$ in (b) and $\theta \simeq {0.40}\pi$ in both plots. The other parameters are the same as in Fig.~\ref{fig2}(a).}
	\label{fig3}
\end{figure}

{The entangled two CMPs can lead to the emission of entangled microwave photons. As analyzed above, the two mechanical sidebands share entanglement due to the scattering of the same phonon mode. It is natural to conjecture that the two output modes near the two sidebands also get quantum correlated, which is confirmed by Fig.~\ref{fig3}. Figure~\ref{fig3}(a) shows the entanglement of the output modes with central frequencies $\Omega_1 = -\omega_b$ and $\Omega_2 = \omega_b$ (in the rotating frame of the drive frequency $\omega_0$) as a function of the inverse bandwidth $\omega_b \tau$. Clearly, the entanglement increases as the output mode bandwidth reduces, because 
the quantum correlation between the two sidebands is established due to the scattering of microwave photons by the mechanical oscillator, which is maximized at frequencies $\omega_0\pm \omega_b$, and a narrower bandwidth more effectively captures this correlation. 
Figure~\ref{fig3}(b) shows that the entanglement reduces as the output modes deviate from the mechanical sidebands. 
Although a larger bandwidth gives rise to smaller entanglement (red line), it shows more resistance to the deviation.  }

\begin{figure}[t]
	\includegraphics[width=\linewidth]{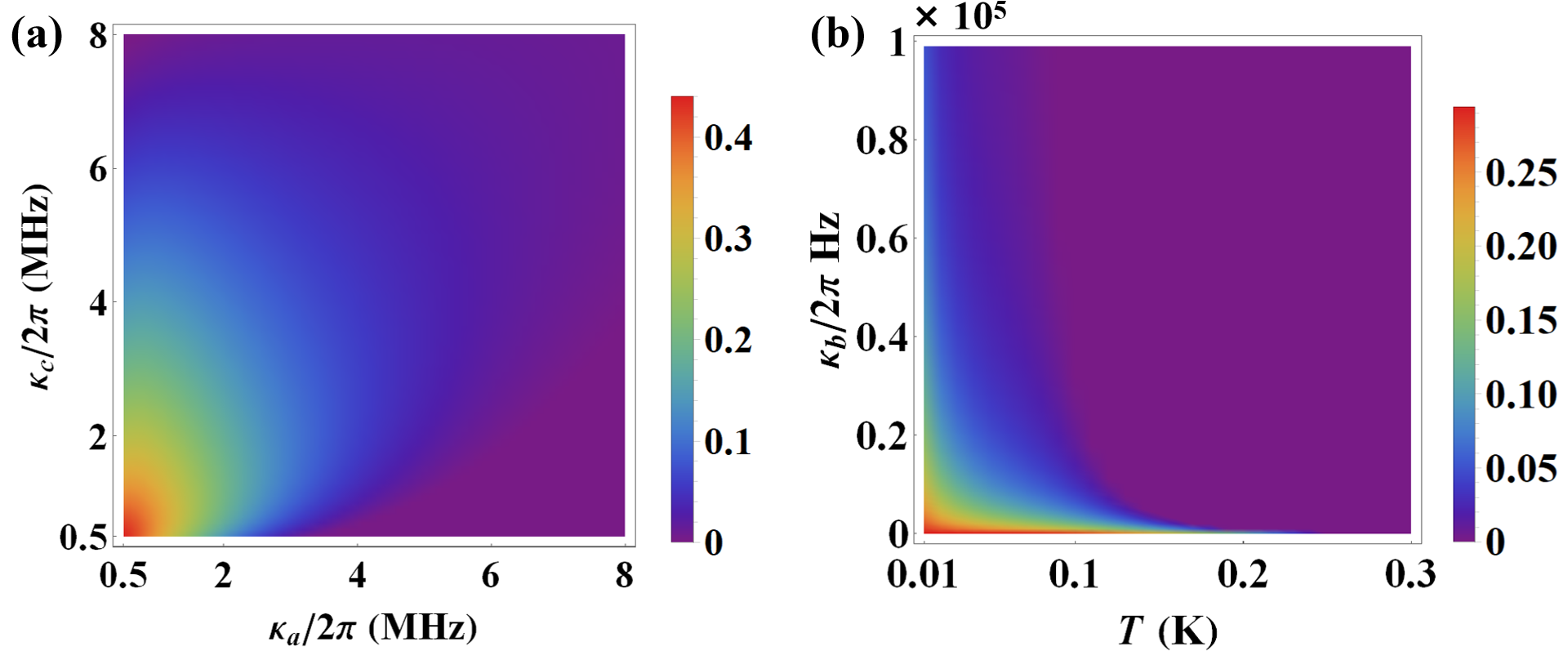}
	\caption{Stationary polariton entanglement $E_N$ versus (a) dissipation rates $\kappa_a$ and $\kappa_c$; (b) bath temperature $T$ and $\kappa_b$. We take  $\theta \simeq {0.40}\pi$, and the other parameters are the same as in Fig.~\ref{fig2}(a).}
	\label{fig4}
\end{figure}

It should be noted that the results of Fig.~\ref{fig2} are obtained under the condition of $\kappa_a \simeq \kappa_c$.  For the case of $\delta \kappa$ being comparable to or larger than $\kappa_{a(c)}$, the associated coupling terms between the two CMPs (cf. Eq.~\eqref{QLEAAfluc}) may have a significant impact on the entanglement.  In Fig.~\ref{fig4}(a), we plot the entanglement versus the two dissipation rates $\kappa_{a}$ and $\kappa_{c}$. It shows that for a wide range of $\kappa_{a(c)}$, the entanglement is present.   The entanglement is also robust with respect to the bath temperature $T$ and the mechanical damping rate $\kappa_b$, as shown in Fig.~\ref{fig4}(b). The entanglement survives for the temperature up to $T \approx {220}$ mK ($\kappa_b$ up to $\sim 2\pi\times10^5$ Hz), under experimentally feasible parameters.

 \section{Conclusion}
 \label{conc}


We present a mechanism of entangling two CMPs in cavity magnonics by dispersively coupling magnons with vibration phonons via introducing magnetostriction. The presence of the vibration phonon mode activates the PDC and the state-swap interactions associated with the two polariton modes, which become entangled when the strengths of the two interactions are properly chosen. 
The work fills the gap in the study of entangling two CMPs.  The entangled CMPs can lead to entangled microwave output fields, which find a wide range of applications in microwave quantum information processing and quantum metrology. The entangled CMPs, as probes, could also improve the detection sensitivity in the dark matter search experiments using CMPs, i.e., ferromagnetic axion haloscopes~\cite{dm1,dm2,dm3}, exploiting the fact that entangled probes can reduce the shot noise and thus increase the detection sensitivity~\cite{Lloyd,Xia,Clerk23}.  

{Our protocol can also be applied to various other polariton systems, e.g., it can be used to entangle exciton polaritons in an exciton-optomechanical system~\cite{PVS}, where excitons and microcavity photons are strongly coupled forming exciton polaritons, and to entangle plasmon polaritons in a plasmon-optomechanical system, where the strong coupling between plasmons and photons leads to plasmon polaritons and the mechanical mode can be the vibration of the molecule trapped in a picocavity~\cite{Benz}.}

\section*{ACKNOWLEDGMENTS}

This work was supported by National Key Research and Development Program of China (Grant No. 2022YFA1405200, 2024YFA1408900), National Natural Science Foundation of China (Grant No. 12474365, 92265202), and Zhejiang Provincial Natural Science Foundation of China (Grant No. LR25A050001).

\section*{APPENDIX A: POLARITON MODES}\label{appA}

\setcounter{figure}{0}
\setcounter{equation}{0}
\setcounter{table}{0}
\renewcommand\theequation{A\arabic{equation}}
\renewcommand\thefigure{A\arabic{figure}}
\renewcommand\thetable{A\arabic{table}}

The Hamiltonian of the cavity and magnon modes with a beam-splitter-type linear coupling reads
\begin{align}
	\begin{split}
		H/\hbar &= \! \omega_{a} a^\dagger a + \omega_{c} c^\dagger c + g \left(a^\dagger c+ac^\dagger \right)\\
	     		 &= 	
	\begin{pmatrix}a^\dagger & c^\dagger\end{pmatrix}
	\begin{pmatrix}\omega_a & g\\g & \omega_c\end{pmatrix}
	\begin{pmatrix}a\\	c\end{pmatrix},
	\end{split}	
\end{align}
where $a$ and $c$ ($[O,O^\dagger]=1$, $O=a,c$) are the annihilation operators of the cavity and magnon modes with resonance frequencies $\omega_a$ and $\omega_c$, and $g$ is the coupling strength. When the two modes are nearly resonant and strongly coupled, $g>\kappa_a, \kappa_c$, with $\kappa_{a}$ and $\kappa_{c}$ being their dissipation rates, the cavity and magnon modes form two CMPs.  By diagonalizing the interaction matrix, we obtain the Hamiltonian of the two CMP modes
\begin{align}
		H/\hbar =  \! \omega_{+} A_+^\dagger A_+ + \omega_{-} A_-^\dagger A_- ,
\end{align}
where $A_+$ and $A_-$ ($[O,O^\dagger]=1$, $O=A_+,\;A_-$) denote the two polariton modes, and $\omega_{+}$ and $\omega_{-}$ are their eigenfrequencies
\begin{align}\label{eigenfreq}
	\begin{split}
	\omega_+&=\frac{\omega_a+\omega_c+\sqrt{(\omega_a-\omega_c)^2+4g^2}}{2},\\
	\omega_-&=\frac{\omega_a+\omega_c-\sqrt{(\omega_a-\omega_c)^2+4g^2}}{2}.
	\end{split}
\end{align}
Clearly, for the resonant case $\omega_a=\omega_c$, $\omega_+-\omega_-=2g$.  The two CMP modes $A_{\pm}$ are the hybridization of the original cavity and magnon modes, reflected in the following transformation matrix
\begin{align}
	\begin{pmatrix}
	A_+\\
	A_-
	\end{pmatrix}=
	\begin{pmatrix}
	\cos\theta & \sin\theta\\
	-\sin\theta & \cos\theta
	\end{pmatrix}
	\begin{pmatrix}
	a\\
	c
	\end{pmatrix},
\end{align}
where \begin{align}\label{tan}
	\begin{split}
	 \theta &=\frac{1}{2} \arctan \frac{2g}{\omega_a-\omega_c}.
	\end{split}
\end{align}

Including vibration phonons, the Hamiltonian of the cavity magnomechanical system under study is given by (i.e., Eq.~\eqref{HHH} in the main text)
\begin{align}\label{HHHs}
	\begin{split}
		H/\hbar &= \! \omega_{a} a^\dagger a + \omega_{c} c^\dagger c + \omega_b b^\dagger b + G_0 c^\dagger c \left( b + b^\dagger \right)		
		\\& + g \left(a^\dagger c+ac^\dagger \right) + i\Omega \left(c^\dagger e^{-i\omega_0t}-c e^{i\omega_0t} \right).
	\end{split}
\end{align}
For strongly coupled cavity and magnon modes, it is more convenient to describe the system with polariton operators $A_{\pm}$. The Hamiltonian~\eqref{HHHs} can then be rewritten as
\begin{align}\label{HHHss}
	\begin{split}
		H/\hbar &= \! \omega_{+} A_+^\dagger A_+ + \omega_{-} A_-^\dagger A_- + \omega_b b^\dagger b		
		\\&+ G_0 \left( b + b^\dagger \right) \Big(A_+^\dagger A_+ \sin^2\theta+ A_+^\dagger A_- \sin\theta \cos\theta 
		\\& \,\,\,\,\,\,\,\,\,\,\,\,\,\,\,\,\,\,\,\,\,\,\,\,\,\,\,\,\,\,\,\,\,\,\, + A_-^\dagger A_+ \cos\theta \sin\theta + A_-^\dagger A_- \cos^2\theta \Big)
		\\&+ i\Omega \Big( A_+^\dagger \sin\theta e^{-i\omega_0t}+A_-^\dagger \cos\theta e^{-i\omega_0t}
		\\&\,\,\,\,\,\,\,\,\,\,\,\,\,\, -A_+ \sin\theta e^{i\omega_0t}-A_- \cos\theta e^{i\omega_0t} \Big) .
	\end{split}
\end{align}
Working in the interaction picture with respect to $\hbar \omega_0 (A_+^\dagger A_+ + A_-^\dagger A_-)$, the Hamiltonian \eqref{HHHss} becomes in the following form
\begin{align}\label{SMHamil}\
	\begin{split}
		H/\hbar &= \! \Delta_{+} A_+^\dagger A_+ + \Delta_{-} A_-^\dagger A_- + \omega_b b^\dagger b  + G_0 \left( b + b^\dagger \right)
		\\& \times \left[A_+^\dagger A_+ \sin^2\theta + A_-^\dagger A_- \cos^2\theta  +  \frac{1}{2} \left(A_+^\dagger A_-  + A_-^\dagger A_+ \right) \sin2\theta \right]
		\\&+ i\Omega \left(A_+^\dagger \sin\theta + A_-^\dagger \cos\theta - A_+ \sin\theta - A_- \cos\theta \right) ,
	\end{split}
\end{align}
where the polariton-drive detunings $\Delta_{\pm} = \omega_{\pm}-\omega_0$.

\section*{APPENDIX B: LINEARIZED QUANTUM LANGEVIN EQUATIONS}
\label{appB}

\setcounter{figure}{0}
\setcounter{equation}{0}
\setcounter{table}{0}
\renewcommand\theequation{B\arabic{equation}}
\renewcommand\thefigure{B\arabic{figure}}
\renewcommand\thetable{B\arabic{table}}

The Hamiltonian~\eqref{HHHs} corresponds to the following quantum Langevin equations (QLEs) by including the dissipation and input noise of each mode:
\begin{align}\label{QLEacb}
	\begin{split}
		\dot{a}=&-(i\Delta_a + \kappa_a) a - i g c + \sqrt[]{2\kappa_a}a^{in},  \\
		\dot{c}=&-(i\Delta_c + \kappa_c) c - i g a - i G_0 c ( b + b^\dagger ) + \Omega + \sqrt[]{2\kappa_c}c^{in}, \\
		\dot{b}=&-i\omega_b b - i G_0 c^\dagger c - \kappa_b b + \sqrt[]{2\kappa_b} b^{in} , \\
	\end{split}
\end{align}
where $\Delta_a=\omega_a-\omega_0$, and $\Delta_c=\omega_c-\omega_0$.   The above QLEs can be rewritten in terms of polariton operators $A_{\pm}$ for strongly coupled cavity and magnon modes, given by 
\begin{align}\label{QLE+-b}
	\begin{split}
		\dot{A}_+=&-i\Delta_+ A_+ - i G_0 \left( b + b^\dagger \right) \left( A_+ \sin^2\theta + A_-\sin \theta \cos \theta \right) 
		\\& - \kappa_+ A_+ - \delta\kappa A_-  + \Omega \sin\theta + \sqrt[]{2\kappa_+}A_+^{in},  \\
		\dot{A}_-=&-i\Delta_- A_- - i G_0 \left( b + b^\dagger \right) \left(A_- \cos^2\theta + A_+\cos \theta \sin \theta \right) 
		\\& - \kappa_- A_- - \delta\kappa A_+ + \Omega \cos\theta + \sqrt[]{2\kappa_-}A_-^{in}, \\
		\dot{b}=&-i\omega_b b - i G_0 \Big(A_+^\dagger A_+ \sin^2\theta+A_+^\dagger A_- \sin\theta \cos\theta 
		\\&+ A_-^\dagger A_+ \cos\theta \sin\theta + A_-^\dagger A_- \cos^2\theta \Big) - \kappa_b b + \sqrt[]{2\kappa_b} b^{in},
	\end{split}
\end{align}
where we define $\kappa_+ \equiv \kappa_a \cos^2\theta + \kappa_c \sin^2\theta $, $\kappa_- \equiv \kappa_a \sin^2\theta + \kappa_c \cos^2\theta $, which can be interpreted as the dissipation rates of the two CMP modes, and $\delta\kappa \equiv (\kappa_c-\kappa_a) \sin\theta \cos\theta$, which denotes the {\it direct} coupling strength between the two CMPs due to the unbalanced dissipation rates $\kappa_a \neq \kappa_c$. The two CMPs can also be {\it indirectly} coupled via the mediation of the phonon mode with the coupling strength of $G_0 {\rm Re} \langle b \rangle \sin 2\theta \equiv {\cal G}$.

Under a strong drive field, the QLEs \eqref{QLE+-b} can be linearized around large average values. We obtain the following linearized QLEs describing the quantum fluctuations of the system $(\delta A_+,\delta A_-, \delta b)$, which are
\begin{align}
	\begin{split}
		\dot{\delta A_+}{=}&\,{-} \big( i\tilde{\Delta}_+ + \kappa_+ \big) \delta A_+ {-} \,\big(i {\cal G}  + \delta\kappa \big) \delta A_- {-} \,G_{+b} \frac{\delta b + \delta b^\dagger}{2} {+}\, \sqrt[]{2\kappa_+}A_+^{in},  \\
		\dot{\delta A_-}{=}&\,{-} \big( i\tilde{\Delta}_- + \kappa_- \big) \delta A_- {-} \,\big(i {\cal G}  + \delta\kappa \big) \delta A_+ {-} \,G_{-b} \frac{\delta b + \delta b^\dagger}{2} {+}\, \sqrt[]{2\kappa_-}A_-^{in},  \\
		\dot{\delta b}\,{=}& \,{-} \big( i \omega_b + \kappa_b \big) \delta b  \,{-} {\left(\frac{G_{+b}}{2} \delta A_+^\dagger + \frac{G_{-b}}{2} \delta A_-^\dagger - \rm{H.\,c.} \right)} + \sqrt[]{2\kappa_b}b^{in} ,	
	\end{split}
\end{align}
where $\tilde{\Delta}_+  = \Delta_+ + 2 G_0 {\rm Re} \langle b \rangle \sin^2\theta$ and $\tilde{\Delta}_-  = \Delta_- + 2 G_0   {\rm Re} \langle b \rangle \cos^2\theta$ are the effective polariton-drive detunings including the frequency shift caused by the dispersive coupling with the phonons. $G_{+b} \equiv G_{+-} \sin \theta$ ($G_{-b}  \equiv G_{+-} \cos \theta$) represents the coupling strength between the polariton $A_+$ ($A_-$) and the phonon mode, where $G_{+-} \equiv G_+ \sin\theta + G_- \cos \theta$, with $G_{\pm}= i 2 G_0\langle A_{\pm} \rangle$ being the enhanced dispersive coupling strengths associated with the two CMPs.

The expressions of the steady-state averages are given by
\begin{align}
	\begin{split}
		\langle A_+ \rangle =& \frac{\Omega}{{\cal D}}  \left[ \delta\kappa  \cos \theta - i \sin\theta \left( \tilde{\Delta}_- - 2 G_0 {\rm Re} \langle b \rangle \cos^2 \theta - i \kappa_- \right) \right] ,\\
		\langle A_- \rangle =& \frac{\Omega}{{\cal D}}  \left[ \delta\kappa \sin \theta - i \cos\theta \left( \tilde{\Delta}_+ - 2 G_0 {\rm Re} \langle b \rangle \sin^2 \theta - i \kappa_+ \right) \right] ,\\
		{\rm Re} \langle b \rangle &=- \frac{G_0}{\omega_b} \big| \langle A_+ \rangle \sin \theta + \langle A_- \rangle \cos \theta \big|^2,
	\end{split}
\end{align}
with $\mathcal{D} = -{\cal G}\, \big( {\cal G} - 2 i \delta\kappa \big)  + \big(\tilde{\Delta}_- - i \kappa_- \big) \big(\tilde{\Delta}_+ - i \kappa_+ \big) + \delta\kappa^2$. The optimal condition for the polariton entanglement corresponds to $\tilde{\Delta}_{+} = -\tilde{\Delta}_{-}  \simeq \omega_b$, as discussed in the main text. Typically, the frequency shift due to the dispersive coupling is much smaller than the resonance frequency $\omega_b$. Therefore, hereinafter we safely take $|\tilde{\Delta}_{\pm}| \simeq |\Delta_{\pm}| \simeq \omega_b$.   This leads to the following simplified expressions
\begin{align}
		\langle A_+ \rangle &\simeq \frac{\delta\kappa \Omega \cos \theta - i \Omega \sin \theta({\Delta}_- - i \kappa_-)}{({\Delta}_- - i \kappa_-)({\Delta}_+ - i \kappa_+) + \delta\kappa^2},\\ \langle A_- \rangle &\simeq \frac{\delta\kappa \Omega \sin \theta - i \Omega \cos \theta({\Delta}_+ - i \kappa_+)}{({\Delta}_- - i \kappa_-)({\Delta}_+ - i \kappa_+) + \delta\kappa^2},
\end{align}
and the QLEs
\begin{align}\label{QLEflucc}
	\begin{split}
		\dot{\delta A_+}{=}&\,{-} \big( i{\Delta}_+ + \kappa_+ \big) \delta A_+ {-}\, \delta\kappa \delta A_- {-} \,G_{+b} \frac{\delta b + \delta b^\dagger}{2} {+}\, \sqrt[]{2\kappa_+}A_+^{in},  \\
		\dot{\delta A_-}{=}&\,{-} \big( i{\Delta}_- + \kappa_- \big) \delta A_- {-} \, \delta\kappa  \delta A_+ {-} \,G_{-b} \frac{\delta b + \delta b^\dagger}{2} {+}\, \sqrt[]{2\kappa_-}A_-^{in},  \\
		\dot{\delta b}\,{=}& \,{-} \big( i \omega_b + \kappa_b \big) \delta b  \,{-} {\left(\frac{G_{+b}}{2} \delta A_+^\dagger + \frac{G_{-b}}{2} \delta A_-^\dagger - \rm{H.\,c.} \right)} + \sqrt[]{2\kappa_b}b^{in} .	
	\end{split}
\end{align}
For simplicity, we neglect the {\it weak} coupling terms ${\cal G} \left(\delta A_+^\dag \delta A_-+ \delta A_+ \delta A_-^\dag \right)$ between the two CMPs, which have a negligible impact on the entanglement. We, however, keep the coupling terms $ -i \delta \kappa \left(\delta A_+^\dag \delta A_-+ \delta A_+ \delta A_-^\dag \right)$, in order to study the impact of a possibly large $\delta \kappa$ of the system on the entanglement.

The QLEs \eqref{QLEflucc} can be expressed in the quadrature form, given by
\begin{align}
	\begin{split}
		\dot{\delta X_+}=&{\Delta}_+ \delta Y_+  - \kappa_+ \delta X_+ - \delta\kappa \delta X_- - {\rm Re} \,G_{+b} \delta X_b + \sqrt[]{2\kappa_+}X_+^{in},  \\
		\dot{\delta Y_+}=&-{\Delta}_+ \delta X_+  - \kappa_+ \delta Y_+ - \delta\kappa \delta Y_- - {\rm Im} \,G_{+b} \delta X_b + \sqrt[]{2\kappa_+}Y_+^{in},  \\
		\dot{\delta X_-}=&{\Delta}_- \delta Y_-  - \kappa_- \delta X_- - \delta\kappa \delta X_+ - {\rm Re} \,G_{-b} \delta X_b + \sqrt[]{2\kappa_-}X_-^{in},  \\
		\dot{\delta Y_-}=&-{\Delta}_- \delta X_-  - \kappa_- \delta Y_- - \delta\kappa \delta Y_+  - {\rm Im} \,G_{-b} \delta X_b + \sqrt[]{2\kappa_+}Y_-^{in},  \\
		\dot{\delta X_b}=&\ \omega_b \delta Y_b - \kappa_b \delta X_b + \sqrt[]{2\kappa_b}X_b^{in},\\
		\dot{\delta Y_b}=& -\omega_b \delta X_b - \kappa_b \delta Y_b  {- {\rm Im} \,G_{+b} \delta X_+} + {\rm Re} \,G_{+b} \delta Y_+ \\& {- {\rm Im} \,G_{-b} \delta X_-} + {\rm Re} \,G_{-b} \delta Y_-  + \sqrt[]{2\kappa_b}Y_b^{in},
	\end{split}
\end{align}
where $\delta X_{\pm}=(\delta A_{\pm} +\delta A_{\pm}^\dagger)/\!\sqrt{2}$, $\delta Y_{\pm}=i(\delta A_{\pm}^\dagger - \delta A_{\pm})/\!\sqrt{2}$, and $\delta X_b=(\delta b + \delta b^\dagger)/\!\sqrt{2}$, $\delta Y_b=i (\delta b^\dagger - \delta b)/\!\sqrt{2}$. The quadratures of the input noises $X_j^{in}$ and $Y_j^{in}$ ($j=+,-,b$) are defined in the same way. The above QLEs can be cast in the simple matrix form
\begin{align}
	\begin{split}
	\dot{u}(t)={\cal R}u(t) + n(t),
	\end{split}
\end{align}
where $u(t)=[\delta X_+(t),\delta Y_+(t),\delta X_-(t),\delta Y_-(t),\delta X_b(t),\delta Y_b(t)]^{\rm T}$, $n(t)=[\sqrt[]{2\kappa_+}X_+^{in},\sqrt[]{2\kappa_+}Y_+^{in},\sqrt[]{2\kappa_-}X_-^{in},\sqrt[]{2\kappa_-}Y_-^{in},\sqrt[]{2\kappa_b}X_b^{in},\sqrt[]{2\kappa_b}Y_b^{in}]^{\rm T}$, and the drift matrix ${\cal R}$ is given by
\begin{align}
	\cal R=\begin{pmatrix}
		-\kappa_+ & {\Delta}_+ & -\delta\kappa & 0 & - {\rm Re}\,G_{+b} &0 \\
		-{\Delta}_+ & -\kappa_+ &0 & -\delta\kappa& - {\rm Im}\,G_{+b} & 0\\
		-\delta\kappa & 0 & -\kappa_- & {\Delta}_- & - {\rm Re}\,G_{-b} &0 \\
		0 & -\delta\kappa&-{\Delta}_- & -\kappa_- & - {\rm Im}\,G_{-b} & 0\\
		0 & 0 & 0 & 0 & -\kappa_b & \omega_b \\
		{- {\rm Im}\,G_{+b}} &{\rm Re}\,G_{+b} & {- {\rm Im}\,G_{-b}} & {\rm Re}\,G_{-b} & -\omega_b & -\kappa_b\\
	\end{pmatrix}.
\end{align}

{
\section*{APPENDIX C: COVARIANCE MATRIX OF TWO OUTPUT MODES}
\label{appC} }

\setcounter{figure}{0}
\setcounter{equation}{0}
\setcounter{table}{0}
\renewcommand\theequation{C\arabic{equation}}
\renewcommand\thefigure{C\arabic{figure}}
\renewcommand\thetable{C\arabic{table}}

{
Here, we show how to define the output modes and obtain their CM. Using the input-output relation, we get the output field of the microwave cavity 
\begin{align}
	\begin{split}
	\delta a^{\rm out}(t) = \sqrt{2 \kappa_a} \delta a(t) - a^{\rm in}(t),
	\end{split}
\end{align}
which is continuous in the frequency spectrum. Therefore, a filter is required to extract independent optical modes at different frequencies, which allows us to define a set of output modes~\cite{Genes08}
\begin{align}
	\begin{split}
	\delta a^{\rm out}_k(t) = \int^t_{-\infty} ds \, g_k (t-s) \,  \delta a^{\rm out}(t),
	\end{split}
\end{align}
where the filter function $g_k(t) = \frac{H(t) - H(t - \tau)}{\sqrt{\tau}}e^{-i \Omega_k t}$ ($k=1,2$), with the Heaviside step function $H$, the filter central frequency $\Omega_k$, and the filter bandwidth $\sim 1/\tau$. }

{
The CM of the magnon, mechanical, and two output modes is defined as
\begin{align}\label{Out}
	\begin{split}
	V_{ij}^{\rm out} = \frac{1}{2} \big\langle u_i^{\rm out}(t)u_j^{\rm out}(t) + u_j^{\rm out}(t)u_i^{\rm out}(t) \big\rangle,
	\end{split}
\end{align}
where
\begin{align}
	\begin{split}
	u^{\rm out}(t)=\big[&\delta X_c(t), \delta Y_c(t), \delta X_b(t), \delta Y_b(t),\\ &\delta X_1^{\rm out}(t), \delta Y_1^{\rm out}(t), \delta X_2^{\rm out}(t), \delta Y_2^{\rm out}(t) \big]^{\rm T}
	\end{split}
\end{align}
is the vector of the quadrature fluctuations of the four modes and the output mode quadratures are defined as $\delta X^{\rm out}_k = (\delta a^{\rm out}_k(t) +\delta a^{\rm out}_k (t)^\dagger)/\!\sqrt{2}$, and $\delta Y^{\rm out}_k = i (\delta a^{\rm out}_k (t)^\dagger - \delta a^{\rm out}_k(t))/\sqrt{2}$. }

{
Taking the Fourier transform of Eq.~\eqref{Out} and incorporating the correlation functions of the input noises, we obtain the steady-state CM
\begin{align}
	\begin{split}
	V^{\rm out} =& \int  d\omega \tilde{T}(\omega) \left( \tilde{M}^{\rm ext}(\omega) + \frac{P^{\rm out}}{2 \kappa_a} \right)\\ \times &\,\, D^{\rm ext} \left( \tilde{M}^{\rm ext}(\omega)^\dagger + \frac{P^{\rm out}}{2 \kappa_a} \right) \tilde{T}(\omega)^\dagger,
	\end{split}
\end{align}
where $P^{\rm out} = \mathrm{Diag}[0,0,0,0,1,1,1,1]$ acts as the projector onto the 4-dimensional space associated with the output quadratures, and
\begin{align}
	\begin{split}
	\tilde{M}^{\rm ext}(\omega) = (i \omega + {\cal R}^{\rm ext})^{-1},
	\end{split}
\end{align}
with
\begin{align}
	{\cal R}^{\rm ext} = \begin{pmatrix}
		-\kappa_c & \Delta_c & -G_{\rm cb} & 0 & 0 & g & 0 & g \\
		-\Delta_c & -\kappa_c & 0 & 0 & -g & 0 & -g & 0 \\
		0 & 0 & -\kappa_b & \omega_b & 0 & 0 & 0 & 0 \\
		0 & G_{\rm cb} & -\omega_b & -\kappa_b & 0 & 0 & 0 & 0 \\
		0 & g & 0 & 0 & -\kappa_a & \Delta_a & 0 & 0 \\
		-g & 0 & 0 & 0 & -\Delta_a & -\kappa_a & 0 & 0 \\
		0 & g & 0 & 0 & 0 & 0 & -\kappa_a & \Delta_a \\
		-g & 0 & 0 & 0 & 0 & 0 & -\Delta_a & -\kappa_a \\
	\end{pmatrix},
\end{align}
which is obtained by linearizing Eq.~\eqref{QLEacb} and writing in the quadrature form. The effective magnomechanical coupling strength $G_{\rm cb} = i 2 G_{0} \langle c \rangle$, where the steady-state average of the magnon mode is given by $\langle c \rangle = \frac{i \Omega \Delta_a}{g^2 - \Delta_c \Delta_a}$. The diffusion matrix $D^{\rm ext}$ reads
\begin{align}
	D^{\rm ext} = \begin{pmatrix}
		\kappa_c & 0 & 0 & 0 & 0 & 0 & 0 & 0 \\
		0 & \kappa_c & 0 & 0 & 0 & 0 & 0 & 0 \\
		0 & 0 & \kappa_b (2 N_b + 1) & 0 & 0 & 0 & 0 & 0 \\
		0 & 0 & 0 & \kappa_b (2 N_b + 1) & 0 & 0 & 0 & 0 \\
		0 & 0 & 0 & 0 & \kappa_a & 0 & \kappa_a & 0 \\
		0 & 0 & 0 & 0 & 0 & \kappa_a & 0 & \kappa_a \\
		0 & 0 & 0 & 0 & \kappa_a & 0 & \kappa_a & 0 \\
		0 & 0 & 0 & 0 & 0 & \kappa_a & 0 & \kappa_a \\
	\end{pmatrix}.
\end{align}
$\tilde{T}(\omega)$ is the Fourier transform of the transformation matrix $T(t)$, which takes the following form~\cite{Genes08}  }
\begin{widetext}
{\begin{align}
	T(t) = \begin{pmatrix}
		\delta(t) & 0 & 0 & 0 & 0 & 0 & 0 & 0 \\
		0 & \delta(t) & 0 & 0 & 0 & 0 & 0 & 0 \\
		0 & 0 & \delta(t) & 0 & 0 & 0 & 0 & 0 \\
		0 & 0 & 0 & \delta(t) & 0 & 0 & 0 & 0 \\
		0 & 0 & 0 & 0 & \sqrt{2\kappa_a}{\rm Re} [g_1(t)] & -\sqrt{2\kappa_a}{\rm Im}[g_1(t)] & 0 & 0 \\
		0 & 0 & 0 & 0 & \sqrt{2\kappa_a}{\rm Im}[g_1(t)] & \sqrt{2\kappa_a}{\rm Re}[g_1(t)] & 0 & 0 \\
		0 & 0 & 0 & 0 & 0 & 0 & \sqrt{2\kappa_a}{\rm Re}[g_2(t)] & -\sqrt{2\kappa_a}{\rm Im}[g_2(t)] \\
		0 & 0 & 0 & 0 & 0 & 0 & \sqrt{2\kappa_a}{\rm Im}[g_2(t)] & \sqrt{2\kappa_a}{\rm Re}[g_2(t)] \\
	\end{pmatrix}.
\end{align}  }
\end{widetext}

\end{document}